# The Robust Orlicz Risk with an Application to the Green Photovoltaic Power Generation

Hidekazu Yoshioka[1,*] and Motoh Tsujimura[2]

*Abstract*—We propose a novel recursive utility for controlling stochastic processes under risk and uncertainty. Our formulation uses a robustified Orlicz risk that can evaluate risk and uncertainty simultaneously. We focus on a control problem of a photovoltaic power generation system that supplies excess electricity to a secondary purpose for generating green hydrogen. The corresponding Hamilton–Jacobi–Bellman equation having a novel nonlinear term is then derived. Computational examples with the available data are finally presented, demonstrating that our methodology can be used for the photovoltaic power generation under different meteorological and operational conditions.

## 1. INTRODUCTION

### 1.1. Background

Green and renewable energy supply has long been a momentous issue toward the sustainable coexistence between the earth environment and human society [1]. There exist diverse renewable energy sources, including but are not limited to hydropower [2], wind power [3], solar power [4], and biomass [5]. They potentially serve as attractive energy sources contributing to the green society, while they depend on the complex and stochastic nature of the environment, and hence there is always a risk of energy shortage.

Stochastic differential equations (SDEs) [6] have been effective mathematical tools for modeling and analysis of renewable energy management. Nonlinear SDEs driven by Brownian motions have been used for efficiently describing the meteorological dynamics, such as the wind speed [7,8] and the cloud cover [9,10]. The models in the literature indeed enable efficiently describing the system dynamics, while their simplicity sometimes suffers from the model uncertainty (i.e., model misspecification) due to the incomplete knowledge. We therefore need to cope with the issues of both risk and uncertainty for achieving the robust supply of green energy in the future. However, mathematical theories for such approaches are still scarce to the best of the authors' knowledge.

### 1.2. Contribution

Based on these research backgrounds, we propose a simple while novel recursive utility for controlled SDEs of the photovoltaic energy management such that both the risk and uncertainty are consistently handled within a unified theory. Our target system is a two-variable SDEs governing the cloud cover dynamics and battery storage. Solutions to the SDEs have bounded ranges due to the degenerate diffusion.

Our contributions are summarized as follows:
- ✓ Proposal of a dynamic recursive utility to evaluate risk and uncertainty.
- ✓ Derivation of the optimality equation associated with the recursive utility.
- ✓ Numerical computation of the recursive utility using real data.

Each contribution is explained as follows.

#### 1.2.1 Recursive Utility

In this study, "risk" means an event where some index (quantity inside the expectation in an objective function) takes a small value. By contrast, "uncertainty" means that parameter values and/or functional shapes in a model are not completely known due to the lack of our knowledge about the physics (cloud cover) of the target dynamics. The robustified Orlicz risk can simultaneously account for both the risk and uncertainty by using an Orlicz function for the former and a Kullback–Leibler divergence for the latter. Our recursive utility uses the robustified dynamic Orlicz risk recently proposed in the insurance [11], which is called the Orlicz risk in this paper. The Orlicz risk resembles a dynamic programming principle thanks to the use of a time-consistent nonlinear expectation suited to stochastic control of diffusion processes. However, its applications have been limited to such as the static optimization problems [12], and more recently, to an environmental restoration problem [13] utilizing a restricted functional form. This paper focuses on a problem of photovoltaic power generation as an application, while our recursive utility generalizes their formulations and can be applied to generic controlled stochastic processes not necessarily

This work was supported in part by KAKENHI No. 22K14441 and 22H02456. (*Corresponding author: H. Yoshioka).

[1] H. Yoshioka is with Japan Advanced Institute of Science and Technology, 1-1 Asahidai, Nomi, Ishikawa 923-1292, Japan (Tel: +81-761-51-1745; e-mail: yoshih@jaist.ac.jp).

[2] Motoh Tsujimura is with Graduate School of Commerce, Doshisha University, Karasuma-Higashi-iru, Imadegawa-dori, Kamigyo-ku, Kyoto 602-8580, Japan (e-mail: mtsujimu@mail.doshisha.ac.jp).

arising in energy management problems.

*1.2.2 Optimality Equation*
We derive an optimality equation associated with the recursive utility to find its optimal control. Indeed, we formally derive Hamilton–Jacobi–Bellman (HJB) equation, which is the optimality equation associated with the recursive utility. We show that it has novel nonlinear terms and a control-dependent discount term, both arising from the robust Orlicz risk. The entropic value-at-risk, a risk measure, has been nested to obtain an HJB equation by properly scaling the risk level [14]. Our approach also employs a scaling of the uncertainty level to obtain an HJB equation. A connection between the Orlicz risk and an auxiliary control problem with an uncertainty-dependent discount factor is discussed as well.

*1.2.3 Computation and Application*
We present a finite difference method for the HJB equation to compute its numerical solutions as well as optimal controls. Our numerical method is explicit in time, and nonlinearity of our HJB equation is handled without resorting to nonlinear iteration methods used in the literature [15,16]. In the finite difference method, the HJB equation is discretized at each space-time point without using any global interpolation functions such as radial basis and neural networks [17,18]. We finally present a computational application of the HJB equation to a photovoltaic power system whose excessive energy can be used for generating green hydrogen contributing to the sustainable development [19,20]. The SDE of the cloud cover dynamics is identified at different study sites in Japan and the corresponding HJB equation is computationally investigated.

*1.2.4 Organization of The Paper*
This rest of the paper is organized as follows. **Section 2** presents the SDE system. **Section 3** presents the recursive utility based on the Orlicz risk and derives the HJB equation. **Section 4** is devoted to the model application. **Section 5** presents the summary and conclusion of this paper. **Appendices** contain proofs of propositions and auxiliary results.

## 2. STOCHASTIC MODEL

*2.1. Cloud Cover*

We use a standard complete probability space $(\Omega, \mathcal{F}, \mathbb{P})$. The cloud cover is the covering ratio of clouds over the sky, which ranges from 0 (cloudless sky) to 1 (fully clouded). We describe the could cover as a continuous-time process $X = (X_t)_{t \geq 0}$ bounded in $[0,1]$ whose Itô's SDE is given by [10]:

$$dX_t = r(a - X_t)dt + \sigma X_t(1 - X_t)dB_t, \ t > 0, \ X_0 \in [0,1], \quad (1)$$

where $r$, $a$, $\sigma$ are positive parameters and $B = (B_t)_{t \geq 0}$ is a standard 1-D Brownian motion. The SDE (1) admits a unique continuous path-wise solution bounded in $[0,1]$ (Yoshioka et al. [10]).

*2.2. Battery Storage and Control*

We consider a solar panel equipped with a battery that can store and discharge the generated energy. The stored energy is a continuous-time process $Y = (Y_t)_{t \geq 0}$ governed by

$$dY_t = (f(t, X_t) - u_t)dt, \ t > 0, \ Y_0 \in [0, \bar{Y}], \quad (2)$$

where $\bar{Y} > 0$ is the battery capacity, $f : [0, +\infty) \times [0,1] \to \mathbb{R}$ represents the unit-time energy storage through the panel receiving the solar irradiance $I = (I_t)_{t \geq 0}$. We assume

$$f(t, X_t) = \varepsilon A I_t (1 - f_0 X_t^{f_1}) \quad (3)$$

with the efficiency constant $\varepsilon \in (0,1)$ (typically around 0.1 [21]), the panel area $A > 0$ and regression constants $f_0 = 0.81$ and $f_1 = 1.9$ [10]. The process $u = (u_t)_{t \geq 0}$ is a control variable of the system representing the discharge for the electricity supply and/or hydrogen generation. This $u$ should satisfy the following condition with which the battery storage is constrained in $[0, \bar{Y}]$ almost surely (a.s.): at each $t > 0$, we require

$$\underline{U}(t, X_t, Y_t) \leq u_t \leq \bar{U}(t, X_t, Y_t), \quad (4)$$

where $\underline{U}$, $\bar{U}$ are given with a constant $U > 0$ as follows:

$$\underline{U} = \begin{cases} f(t, X_t) & (Y_t = \bar{Y}) \\ 0 & (Y_t < \bar{Y}) \end{cases}, \quad \bar{U} = \begin{cases} 0 & (Y_t = 0) \\ U & (0 < Y_t < \bar{Y}) \\ \max\{U, f(t, X_t)\} & (Y_t = \bar{Y}) \end{cases}. \quad (5)$$

We assume that there is a target process $\lambda = (\lambda_t)_{t \geq 0}$ bounded in the range $[0, U]$ such that the discharge should meet the target ($u_t = \lambda_t$) and the residual $U - u_t \geq 0$ can be used for generating the green hydrogen. If the stored energy level is sufficiently high, then the controls should always be $u_t = \lambda_t$, while it is not always possible to achieve this equality because the solar power is stochastic, and the battery capacity is finite. Finally, the collection of measurable processes $u$ satisfying (4) and (5) is denoted as $\mathcal{A}$.

### 2.3. Model Uncertainty

Model uncertainty is represented by a distortion of the drift of the SDE (1) using the robust control approach of Hansen-Sargent [22]. Set the drifted Brownian motion $W = (W_t)_{t \geq 0}$ by

$$dW_t = dB_t - \phi_t dt, \quad (6)$$

where $\phi = (\phi_t)_{t \geq 0}$ is a real-valued measurable process representing the model uncertainty. The probability measure under which $W$ becomes a standard Brownian motion is denoted as $\mathbb{Q}(\phi)$, which serves as a distorted probability measure considering model uncertainty as explained in the next section. The existence of $\mathbb{Q}(\phi)$ is guaranteed if the following condition is satisfied [22]:

$$\mathbb{E}_\mathbb{P}\left[\exp\left(\int_0^t \phi_s^2 ds / 2\right)\right] < +\infty \text{ for each } t > 0. \quad (7)$$

Here, an expectation on a generic probability measure $\mathbb{Q}$ is denoted as $\mathbb{E}_\mathbb{Q}$. The collection of real-valued measurable processes $\phi = (\phi_t)_{t \geq 0}$ satisfying (7) is denoted as $\mathcal{B}$. The unit-time relative entropy $c(\phi)$, the Kullback–Leibler divergence, measuring the difference between $\mathbb{P}$ and $\mathbb{Q}(\phi)$ at $t$ is calculated as $\phi_t^2 / 2$ [22].

**Remark 1** We use the relative entropy as a measure of the difference between two probability measures $\mathbb{P}$ and $\mathbb{Q}(\phi)$, while more generalized one such as the Tsallis and Rényi divergences [23] may also be used if necessary.

### 3. ORLICZ RISK AND HJB EQUATION

### 3.1. Recursive Utility

The recursive utility as the worst-case optimized objective through operating the system is introduced by generalizing the previous ones [11,13]. We set a smooth, increasing, and convex function $\Phi : [0, +\infty) \to [0, +\infty)$ (Orlicz function) such that $\Phi(0) = 0$ and $\Phi(1) = 1$. We set a coefficient $C(\phi) \geq 0$ measuring difference between $\mathbb{P}$ and $\mathbb{Q}(\phi)$ such that $C(\phi)|_{\phi \equiv 0} = 0$ and is $\mathcal{F}_t$-measurable. According to Bellini et al. [11], at time $t \geq 0$, the Orlicz risk $\|Z\|_{\Phi,t}$ of a random variable $Z$ that is positive a.s. $\mathcal{F}_t$-measurable is given by

$$\|Z\|_{\Phi,t} = \inf\left\{h > 0 \text{ is } \mathcal{F}_t\text{-measurable} \,\Big|\, \sup_{\mathbb{Q}(\phi)} \mathbb{E}_{\mathbb{Q}(\phi)}\left[\underbrace{\Phi\left(\frac{Z}{h}\right)}_{\text{Risk}} - \underbrace{C(\phi)}_{\text{Uncertainty}} \,\Big|\, \mathcal{F}_t\right] \leq 1\right\}. \quad (8)$$

The Orlicz risk (8) contains the two terms in the conditional expectation $\mathbb{E}_{\mathbb{Q}(\phi)}$. The first term represents the risk aversion in a way that a sharper profile $\Phi$ of corresponds to a stronger risk aversion. A major choice would be

$$\Phi(z) = z^p \text{ and } \Phi(z) = \frac{e^{\mu z} - 1}{e^\mu - 1}, \quad z > 0 \quad (9)$$

with risk-aversion strengths $p > 1$ and $\mu > 0$. The left one of (9) has been considered in Yoshioka et al. [13], while we will cover a generic $\Phi$ covering both of (9). The first- and second- derivatives of $\Phi$ are denoted as $\Phi'$, $\Phi''$. The term $C(\phi)$ in combined with "sup" penalizes the uncertainty [11].

We consider a recursive utility $\Psi = (\Psi_t)_{t \geq 0}$ for optimizing the system by invoking the time consistency of the Orlicz risk. We then set a unit-time disutility coefficient $D : [0,T) \times [0,U] \times [0,\bar{Y}] \to [0,+\infty)$. This $D$ also evaluates the hydrogen generation. The non-negativity of $D$ is a key to well-pose the Orlicz risk (i.e., Eq. (11)). The coefficient $D$ is set as

$$D(t, u_t, Y_t) = \frac{(\lambda_t - u_t)_+^2}{2} + w_1 \frac{(U - u_t)_+^2}{2} + w_2 \mathbb{I}(Y_t = 0) \qquad (10)$$

with weights $w_1$, $w_2 > 0$ and $(\cdot)_+ = \max\{\cdot, 0\}$, where the first term penalizes the supply deficit, the second term the situation where the discharge is not fully utilized where the incentive for generating the hydrogen is evaluated through a penalization of the positive residual $U - u_t$, and the third term the energy shortage with the indicator function $\mathbb{I}(Y_t = 0)$, which is 1 if $Y_t = 0$ and is 0 otherwise.

In time interval $(t, t+k)$ with a small $k > 0$, we set a recursive (dis-)utility process $\Psi$ satisfying the backward recursion:

$$\inf_{u \in \mathcal{A}} \sup_{\phi \in \mathcal{B}} \mathbb{E}_{\mathbb{Q}(\phi)} \left[ \Phi\left( \frac{\Psi_{t+k} + \int_t^{t+k} D(s, u_s, Y_s) \, ds}{\Psi_t} \right) - \int_t^{t+k} \frac{\phi_s^2}{2\eta} ds \, \bigg| \, \mathcal{F}_t \right] = 1 \qquad (11)$$

with the uncertainty-aversion coefficient $\eta > 0$ such that the model uncertainty is more strongly penalized (i.e., operator of the system assumes a larger model uncertainty) by choosing a larger $\eta$. We need to complement (11) by a terminal condition $\Psi_T$ at a terminal time $T > 0$ for its well-posedness. For simplicity, we assume $\Psi_T = 0$ (no penalty at $t = T$).

The recursion (11) has not been found in the literature, and therefore is a novel formulation of the risk-sensitive stochastic control under both model uncertainty.

### 3.2. HJB Equation

The HJB equation corresponding to (11) is presented. Its derivation procedure is shown in **Appendix A**. Set the infinitesimal generator $\mathcal{L}_{u,\varphi,t,x,y}$ for generic smooth $F(t,x,y) : [0,T] \times [0,1] \times [0,\bar{Y}] \to \mathbb{R}$ with $(v, \varphi) \in \mathbb{R}^2$ [6]:

$$\mathcal{L}_{v,\varphi,t,x,y} F = \frac{\partial F}{\partial t} + (r(a-x) + \sigma x(1-x)\varphi) \frac{\partial F}{\partial x} + \frac{\sigma^2 x^2 (1-x)^2}{2} \frac{\partial^2 F}{\partial x^2} + (f(t,x) - v) \frac{\partial F}{\partial y}. \qquad (12)$$

Assume that, with an abuse of notations, the recursive utility $\Psi_t$ has a Markovian form $\Psi_t = \Psi(t, X_t, Y_t)$. Then, we obtain **Proposition 1** below (Proof is in **Appendix A**). We set $\Omega = [0,T) \times [0,1] \times [0,\bar{Y}]$ and collection of all functions continuously differentiable once, twice, and once with respect to the first, second, and third arguments in $\Omega$ is denoted as $C^{1,2,1}(\Omega)$. **Proposition 1** states that $\Psi$ becomes a classical solution (a solution that satisfies an equation at each point in a domain) (e.g., Chapter 6 in Hernández-Lerma et al. [24]) to an HJB equation.

**Proposition 1** Assume that $\Psi \in C^{1,2,1}(\Omega)$. Then, the HJB equation for (11) is obtained in $\Omega$ as

$$\inf_{v \in [\underline{U}(t,x,y), \bar{U}(t,x,y)]} \sup_{\varphi \in \mathbb{R}} \left\{ \begin{array}{l} \mathcal{L}_{v,\varphi,t,x,y} \Psi + D(t,v,y) - \dfrac{\varphi^2}{2\Phi'(1)\eta} \Psi \\ + \dfrac{\Phi''(1)}{2\Phi'(1)\Psi} \left( \sigma x(1-x) \dfrac{\partial \Psi}{\partial x} \right)^2 \end{array} \right\} = 0. \qquad (13)$$

The third and fourth terms of the equation (13) are non-standard. The third term being a formal discount term due to the uncertainty aversion, with the discount rate proportional to the relative entropy. The fourth term comes from the risk aversion due to using the strictly convex $\Phi$. Indeed, it vanishes when $\Phi = x$ as in the classical dynamic programming principle where $\Phi''(1) = 0$. More importantly, the influences of $\Phi$ enter the HJB equation only through $\Phi'(1)$ and $\Phi''(1)$ but not whole the $\Phi$ profile. The behavior of $\Phi$ at the point 1 is therefore important. Note that a solution-dependent uncertainty-aversion parameter as in the last term in the first line of (13) has been employed heuristically (e.g., in economics [25,26]). The HJB equation (13) should be satisfied both inside and along the boundary of $\Omega$, which can be understood in the sense of state-constraint viscosity solutions (See **Appendix B**).

We can also obtain the optimal control $u^* \in \mathcal{A}$ (maximizer) and the corresponding worst-case distortion $\phi^* \in \mathcal{B}$ (minimizer) of (13) as follows:

$$\phi^*(t, X_t, Y_t) = \frac{\Phi'(1)\eta\sigma x(1-x)}{\Psi}\frac{\partial \Psi}{\partial x}, \tag{14}$$

$$u^*(t, X_t, Y_t) = \underset{v \in [\underline{U}(t,X_t,Y_t), \bar{U}(t,X_t,Y_t)]}{\arg\min} \left\{ \mathcal{L}_{v,\phi^*,t,x,y}\Psi + D(t,v,y) \right\}. \tag{15}$$

The right-hand sides of (14)-(15) are evaluated at $(t, X_t, Y_t)$. In this view, finding the recursive utility and the associated optimal control as well as the worst-case model uncertainty reduces to the resolution of the HJB equation (13).

This paper gives another, novel view of the solution- and hence state-dependence by the Orlicz risk. The HJB equation (13) satisfies the so-called Isaacs condition (i.e., the order of "inf" and "sup" is exchangeable) and is rewritten as

$$\begin{aligned}&\frac{\partial \Psi}{\partial t} + r(a-x)\frac{\partial \Psi}{\partial x} + \frac{\sigma^2 x^2 (1-x)^2}{2}\frac{\partial^2 \Psi}{\partial x^2} \\ &+ \underset{v \in [\underline{U}(t,x,y), \bar{U}(t,x,y)]}{\inf}\left\{ (f(t,x)-v)\frac{\partial \Psi}{\partial y} + D(t,v,y) \right\} \\ &+ \frac{1}{2\Psi}\frac{\Phi'(1)^2 \eta + \Phi''(1)}{\Phi'(1)}\left(\sigma x(1-x)\frac{\partial \Psi}{\partial x}\right)^2 = 0 \end{aligned} \tag{16}$$

This is the HJB equation with the net uncertainty-aversion parameter $\eta' = \frac{\Phi'(1)^2 \eta + \Phi''(1)}{\Phi'(1)} > 0$ and the Orlicz function chosen by an identity map. This HJB equation is also the optimality equation of the stochastic differential game with a control-dependent discount rate, i.e., we obtain the proposition below (Its proof is in **Appendix C**).

**Proposition 2** *Under the assumption of **Proposition 1**, (13) is the optimality equation of the control problem; for $0 \leq t \leq T$,*

$$\Psi(t, X_t, Y_t) = \inf_{u \in \mathcal{A}} \sup_{\phi \in \mathcal{B}} \mathbb{E}_{\mathbb{Q}(\phi)}\left[ \int_t^T e^{-\int_t^s \frac{\phi_\tau^2}{2\eta'}d\tau} D(s, u_s, Y_s)ds \bigg| \mathcal{F}_t \right]. \tag{17}$$

The right-hand side of (17) with $\phi \equiv 0$ gives a common objective function of a control problem without any risks and uncertainties. In this view, our recursive utility certainly accounts for the risk and uncertainty.

**Remark 2** For some diffusion process, the HJB equation (13) is solvable exactly. See **Appendix D**.

### 3.3. Finite Difference Method

A fully-explicit finite difference method combining central and one-sided differences is used to discretize the HJB equation. The domain $\Omega$ is discretized by structured vertices $P_{i,j,k} = (\Delta t i, \Delta x j, \Delta y k)$ for $i = 0, 1, 2, ..N_t$, $j = 0, 1, 2, ..N_x$, $k = 0, 1, 2, ..N_y$ ($\Delta t = T/N_t$, $\Delta x = 1/N_x$, $\Delta y = \bar{Y}/N_y$) with some $N_t, N_x, N_y \in \mathbb{N}$. The discretized $\Psi$ at $P_{i,j,k}$ is denoted as $\Psi_{i,j,k}$. Set $t_i = \Delta t i$, $x_j = \Delta x j$, $y_k = \Delta y k$.

Fix a vertex $P_{i,j,k}$ ($i < N_t$). We set

$$p_L = (\Psi_{i+1,j,k} - \Psi_{i+1,j-1,k})/\Delta x, \quad p_R = (\Psi_{i+1,j+1,k} - \Psi_{i+1,j,k})/\Delta x, \tag{18}$$

$$p_D = (\Psi_{i+1,j,k} - \Psi_{i+1,j,k-1})/\Delta y, \quad p_U = (\Psi_{i+1,j,k+1} - \Psi_{i+1,j,k})/\Delta y. \tag{19}$$

Then, we set the discretized HJB equation at $P_{i,j,k}$ as follows:

$$\Psi_{i,j,k} = \Psi_{i+1,j,k} + \Delta t (I_1 + I_2 + I_3), \tag{20}$$

$$I_1 = r(a - x_j) \begin{cases} p_R & (a \geq x_j) \\ p_L & (a < x_j) \end{cases} + \frac{(\sigma x_j (1 - x_j))^2}{2} \frac{p_R - p_L}{\Delta x}, \tag{21}$$

$$I_2 = f(t_i, x_j) \begin{cases} p_U & (j < N_y) \\ p_D & (j = N_y) \end{cases} + \min_{v \in [\underline{U}(t_i, x_j, y_k), \overline{U}(t_i, x_j, y_k)]} \{-vp' + D(x_j, v, y_k)\}, \tag{22}$$

with $p' = p_U$ if $k = 0$ and $p' = p_D$ otherwise,

$$I_3 = \eta' \frac{(\sigma x_j (1 - x_j))^2}{2(\Psi_{i+1,j,k} + \varepsilon)} \overline{p}^2, \tag{23}$$

where $\varepsilon = 10^{-10}$ avoids the possible division by 0, and $\overline{p}^2$ is evaluated by the Godunov-like discretization [27]:

$$\overline{p}^2 = \begin{cases} p_R^2 & (j = 0) \\ \min\{|p_R|, |p_L|\}^2 & (0 < j < N_x, p_L \leq p_R, p_L p_R \geq 0) \\ 0 & (0 < j < N_x, p_L \leq p_R, p_L p_R < 0) \\ \max\{|p_R|, |p_L|\}^2 & (0 < j < N_x, p_L > p_R) \\ p_L^2 & (j = N_x) \end{cases}. \tag{24}$$

The discretized HJB equation (20) is uniquely solvable backward in time without resorting to any matrix solvers. We empirically found that the computed $\Psi$ at each vertex remains positive as long as $\Delta t$ is chosen to be sufficiently small for each fixed $\Delta x$ and $\Delta y$ (See **Appendix E**). Finally, the worst-case model uncertainty (14) and optimal control (15) at each $P_{i,j,k}$ are computed as (14) with $\overline{p}$ considering (24) and the minimizer of (22), respectively.

## 4. APPLICATION

### 4.1. Study Sites

The parameters of the cloud cover dynamics (1) are fitted against the real data available at the two weather observation stations of Japan Meteorological Agency (Kyoto: 35°0.8'N 135°43.9'E, 41 m above sea level and Kanazawa: 36°35.3'N 136°38.0'E, 6 m above sea level). The daily cloud cover time series is available at Japan Meteorological Agency until the end of 2019 (https://www.data.jma.go.jp/gmd/risk/obsdl/). We identify the SDE (1) from the ten-year time series from January 1 2010 to December 31 2019 with no missing value.

The least-squares method based on the forward Kolmogorov equation [10] was used to identify parameters of the SDE (1) at each site as shown in **Table 1**. **Fig. 1** shows that the empirical and fitted stationary probability density functions (PDFs) of $X$ agree well with each other at each site. The PDFs of the fitted models are bimodal for both sites, and that of Kanazawa is more weighted on the right-side extrema corresponding to the cloudy weather. The irradiance is computed according to Appendix B of Yoshioka et al. [10].

**TABLE I**
PARAMETERS FOR THE CLOUD COVER DYNAMICS.

|  | $r$ (1/day) | $a$ (-) | $\sigma$ (1/day$^{1/2}$) |
|---|---|---|---|
| Kyoto | 0.602 | 0.709 | 2.04 |
| Kanazawa | 0.580 | 0.766 | 2.27 |

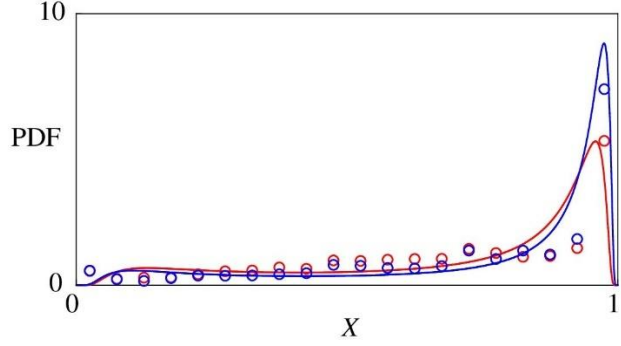

**Fig. 1.** Empirical (circles) and fitted PDFs (curves) of $X$: Kyoto (red) and Kanazawa (blue).

*4.2. Computational Condition*

We consider the system for Kanazawa and Kyoto, assuming that the solar panel is oriented to the south with the surface slope 45 (deg). Unless otherwise specified, we use the non-dimensionalization $\bar{Y}=1$ and $\varepsilon A=0.001$, $\eta=0.1$, $w_1=0.1$, $w_2=0.5$, $\lambda_t \equiv 0.05$ (constant for simplicity), $U=0.2$, $\Phi(z)=z^{3/2}$. The resolution is $N_x = N_y = 300$, $N_t$ is case dependent with $\Delta t = 1/(24\cdot 25 \cdot 60)$ (day). We report that in all the case below we numerically have $\Psi > 0$. See **Appendix E** for auxiliary results. The computational time for the case of Kyoto for one computational year was $1.429\times 10^5$ (s) (a laptop, 12th Gen Intel(R) Core(TM) i7-1260P, 2.1Ghz). The seemingly long computational time is due to using an explicit discretization for the diffusion term without any parallelization. Notice that an objective of this paper is not a comparison of numerical methods, but development of an implementable scheme.

*4.3. On the Nonlinear Term*

We firstly analyze the role played by the unique term, which corresponds to the last term in (13) ($I_3$ in (20)). This term is non-negative, and therefore potentially increases the recursive (dis-)utility $\Psi$. We study the size of its increase by computing the corresponding term at a fixed point $(x,y)=(1/2,1/2)$ during $[0,T)$ with $T=365$ (day), where $t=0$ is January 1 00:00:00. **Fig. 2** shows the computed history of $I_3$, $\Psi$, and $\bar{p}^2$ of (24). The results suggest that the term $I_3$ (in view of a time-backward manner) initially increases, but eventually decays to 0 as the time elapses due to the increase of the value of $\Psi$. This implies that the proposed Orlicz risk is suited to the analysis with a short horizon: at most a weekly scale in the present case. Indeed, in the proposed Orlicz risk, the uncertainty-aversion is understood through a formal discount factor in (17).

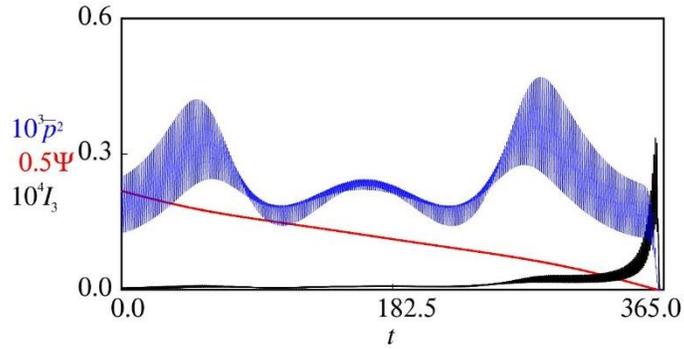

**Fig. 2.** Computed $\bar{p}^2$, $\Psi$, and $I_3$ for Kanazawa.

*4.4. Demonstrative Computational Examples*

We show demonstrative computational examples in this section. We firstly analyze the optimal control in a weekly scale by setting $T=365$ (day). We analyze few weeks containing the terminal time $t=T$ (the end of December 31), which includes the winter solstice (December 22, 355 (day) to 356 (day)) at which the length of the daytime is the shortest in a year. **Figs. 3-4** show the computed $\Psi$ at selected instances near the winter solstice for Kyoto and Kanazawa, respectively. Discharging the stored energy in the battery should be suppressed if the stored energy is low around the winter solstice due to the low irradiance. The operation of the system is slightly more conservative for Kanazawa than Kyoto due to the lower irradiance.

Hereafter, we focus on Kanazawa for simplicity, while qualitatively the same conclusion also applies to Kyoto. **Fig. 5** shows the computed optimal control $u^*$ at selected instances around the summer solstice (June 21, 171 (day) to 172 (day)) where the terminal time $T$ equals 180 (day). The figure demonstrates that the stored energy is discharged more actively around the summer solstice than around the winter solstice of **Fig. 3**.

**Figs. 6-7** compare the optimal residual discharge $R^* = (u^* - \lambda)_+$ that can be used for generating the green hydrogen, for different values of $\eta$ (hence the net uncertainty aversion $\eta'$) and weights $w_1, w_2$, respectively. **Fig. 6** shows that increasing $\eta'$, namely increasing the risk- or uncertainty-aversion through $\Phi''(1)$ or $\eta$, does not qualitatively affect the optimal control unless it is excessively large. The optimal control is robust against the uncertainty from the standpoint of the hydrogen generation. **Fig. 7** shows that increasing the weight $w_1$ concerning the hydrogen generation recommends more actively generating the hydrogen as well. Increasing the weight $w_2$ penalizing the storage depletion by contrast leads to that maintaining the stored energy at an elevated level with less generating the hydrogen under a cloudless sky condition is optimal. As demonstrated in this paper, the proposed model is applicable to the optimization of the system under a variety of conditions.

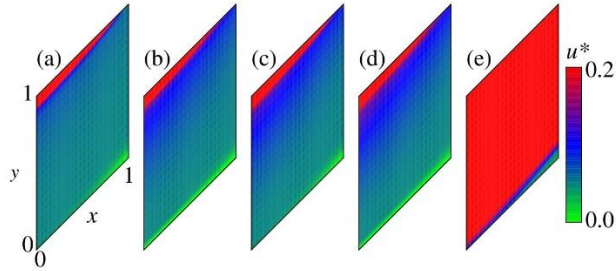

**Fig. 3.** $u^*$ for Kyoto at time (day): (a) 351.5, (b) 355.0, (c) 355.5, (d) 356.0, (e) 364.5.

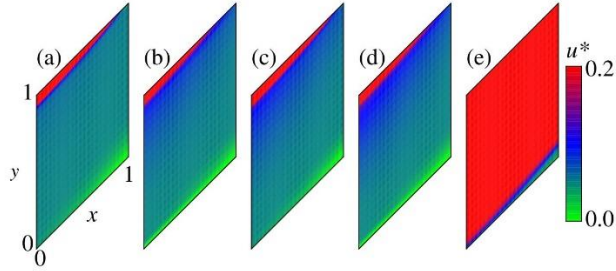

**Fig. 4.** $u^*$ for Kanazawa: the same legends with Fig. 3.

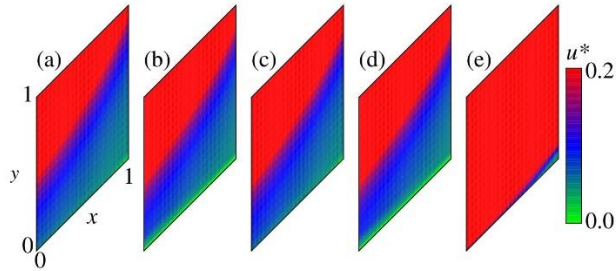

**Fig. 5.** $u^*$ for Kanazawa in summer at time (day): (a) 167.5, (b) 171.0, (c) 171.5, (d) 172.0, (e) 179.5.

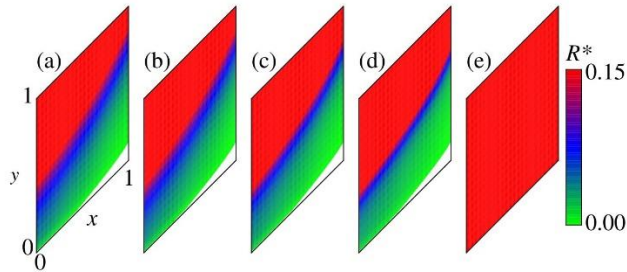

**Fig. 6.** $R^*$ at 171.5 (day). $\eta$ is: (a) 0.01, (b) 0.1, (c) 0.5, (d) 1, (e) 3. No hydrogen generation in the white area.

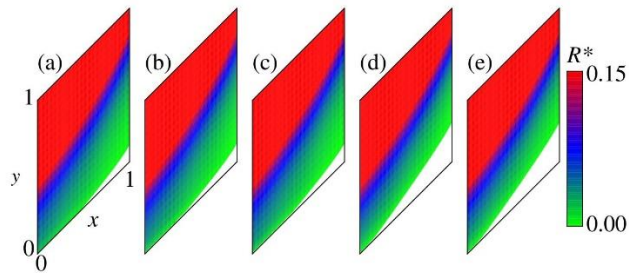

**Fig. 7.** $R^*$ at 171.5 (day). $(w_1, w_2)$ is: (a) $(0.1, 0)$, (b) $(0.1, 0.5)$, (c) $(0.5, 0.5)$, (d) $(0.1, 2.5)$, (e) $(0.5, 2.5)$. No hydrogen generation in the white area.

## 5. CONCLUSION

We proposed a recursive utility based on the robustified dynamic Orlicz risk and derived the associated HJB equation having the novel nonlinear term. Computational applications were demonstrated for the photovoltaic power system whose excess energy can be used for generating the green hydrogen.

We focused only on the application to the Orlicz risk to diffusion processes, while jump-diffusion processes can also be considered with a proper modification. However, the existence of solutions to the HJB equation will become a theoretical issue especially if the jumps have infinite activities as in some applied problems [28]. Adaptation of the proposed approach to mean field models is also of interest from both theoretical and practical standpoints [29-30]. The formalism based on the HJB equation can be incorporated into an integrated model for energy management of real-scale buildings as in Michailidis et al. [31]. The development of a more efficient numerical method will also be an important research topic to be addressed in the future. Investigation of energy management for jump-driven cases will become possible based on the recent work [32].

## REFERENCES

[1] Z. Ahmed *et al.*: *How do green energy technology investments, technological innovation, and trade globalization enhance green energy supply and stimulate environmental sustainability in the G7 countries?*, Gondwana Research, **112**(2022), pp. 105–115. doi: 10.1016/j.gr.2022.09.014

[2] W. Tobias et al., *The future of Alpine Run-of-River hydropower production: Climate change, environmental flow requirements, and technical production potential*, Science of the Total Environment, **890**(2023), 163934. doi: 10.1016/j.scitotenv.2023.163934

[3] A. Bensoussan, B. Chevalier-Roignant, A. Rivera, *A model for wind farm management with option interactions,* Production and Operations Management, **31**(2022), pp. 2853–2871. doi: 10.1111/poms.13721

[4] D. Yang *et al.*, *A review of solar forecasting, its dependence on atmospheric sciences and implications for grid integration: Towards carbon neutrality*, Renewable and Sustainable Energy Reviews, **161**(2022), 112348. doi: 10.1016/j.rser.2022.112348

[5] H. Jahani, H. Gholizadeh, Z. Hayati, H. Fazlollahtabar, *Investment risk assessment of the biomass-to-energy supply chain using system dynamics*, Renewable Energy, **203**(2023), pp. 554–567. doi: 10.1016/j.renene.2022.12.038

[6] B. Øksendal, A. Sulem, Applied Stochastic Control of Jump Diffusions. Springer, Cham, 2019.

[7] G. M. Jónsdóttir, F. Milano, *Data-based continuous wind speed models with arbitrary probability distribution and autocorrelation*, Renewable. Energy, **143**(2019), pp. 368–376. doi: 10.1016/j.renene.2019.04.158

[8] M. L. Sørensen *et al.*, *Recent developments in multivariate wind and solar power forecasting*, Wiley Interdisciplinary Reviews: Energy and Environment, **12**(2023), e465. doi: 10.1002/wene.465

[9] C. A. Thilker, H. Madsen, J. B. Jørgensen, *Advanced forecasting and disturbance modelling for model predictive control of smart energy systems*, Applied Energy, **292**(2021), 116889. doi: 10.1016/j.apenergy.2021.116889

[10] H. Yoshioka, Z. Li, A. Yano, *An optimal switching approach toward cost-effective control of a stand-alone photovoltaic panel system under stochastic environment*, Applied Stochastic Models in Business and Industry, **35**(2019), pp. 1366–1389. doi: 10.1002/asmb.2485

[11] F. Bellini, R. J. Laeven, E. R. Gianin, *Dynamic robust Orlicz premia and Haezendonck–Goovaerts risk measures*, European. Journal of Operations Research, **291**(2021), pp. 438–446. doi: 10.1016/j.ejor.2019.08.049

[12] F. Bellini, R. J. Laeven, E. Rosazza Gianin, *Robust return risk measures*, *Mathematical Finance and Economics*, **12**(2018), pp. 5–32. doi: 10.1007/s11579-017-0188-x

[13] H. Yoshioka, M. Tsujimura, F. Aranishi, T. Tanaka, *Environmental management and restoration under risk and uncertainty unified under a robustified dynamic Orlicz risk*, 2023, Preprint available at http://arxiv.org/abs/2306.01998

[14] A. Pichler, R. Schlotter, *Risk-Averse Optimal Control in Continuous Time by Nesting Risk Measures*, Mathematics of Operations Research, Published online. 2022. doi: 10.1287/moor.2022.1314

[15] W. Qiu, Q. Song, G. Yin, Solving elliptic Hamilton-Jacobi-Bellman equations in a value space. IEEE Control Systems Letters, 5(2020), 55-60. doi: 10.1109/LCSYS.2020.2999650
[16] H. Wu et al., Differential game approach for attack-defense strategy analysis in Internet of Things networks. IEEE Internet of Things Journal, 9(2023), 10340-10353. doi: 10.1109/JIOT.2021.3122115
[17] Y. Qiu et al., Observer-based robust optimal control for helicopter with uncertainties and disturbances. Asian Journal of Control, 25(2023), 3920-3932. doi: 10.1002/asjc.3079
[18] M.L. Greene, Z.I., Bell, S. Nivison, W.E. Dixon, Deep neural network-based approximate optimal tracking for unknown nonlinear systems. IEEE Transactions on Automatic Control, 68(2023), 3171-3177. doi: 10.1109/TAC.2023.3246761
[19] J. E. Fokkema *et al.*, *Seasonal hydrogen storage decisions under constrained electricity distribution capacity*, Renewable Energy, **195**(2022), pp. 76–91. doi: 10.1016/j.renene.2022.05.170
[20] M. B. Bertagni, S. W. Pacala, F. Paulot, A. Porporato, *Risk of the hydrogen economy for atmospheric methane*, Nature communications, **13**(2022), 7706. doi: 10.1038/s41467-022-35419-7
[21] A. Sharma, V. K. Bajpai, Enhancement of power generation efficiency of PV system using mirror reflector, *International Journal of Ambient Energy*, **43**(2022), pp. 8036–8045. doi: 10.1080/01430750.2022.2088615
[22] L. P. Hansen, T. J. Sargent, *Robust control and model uncertainty*, American Economics Review, **91**(2001), pp. 60–66. doi: 10.1257/aer.91.2.60
[23] N. Sajid *et al.*, *Bayesian brains and the Rényi divergence*, Neural Computation, **34**(2022), pp. 829–855. doi: 10.1162/neco_a_01484
[24] O. Hernández-Lerma, L.R. Laura-Guarachi, S. Mendoza-Palacios, D. González-Sánchez An Introduction to Optimal Control Theory: The Dynamic Programming Approach Springer, Cham, 2023.
[25] B. Liu, H. Meng, M. Zhou, *Optimal investment and reinsurance policies for an insurer with ambiguity aversion*, The North American Journal of Economics and Finance, **55**(2021), 101303. doi: 10.1016/j.najef.2020.101303
[26] J. Ma, Y. Wu, *Robust investment and hedging policy with limited commitment*, Economic Modeling, **125**(2023), 106344. doi: 10.1016/j.econmod.2023.106344
[27] S. Osher, C. W. Shu, *High-order essentially nonoscillatory schemes for Hamilton–Jacobi equations*, SIAM Journal of Numerical Analysis, **28**(1991), pp. 907–922. doi: 10.1137/0728049
[28] Y. He, R. Kawai, Y. Shimizu, K. Yamazaki, *The Gerber-Shiu discounted penalty function: A review from practical perspectives*, Insurance: Mathematics and Economics, **109**(2023), pp. 1-28. doi: 10.1016/j.insmatheco.2022.12.003
[29] A. V. Shrivats, D. Firoozi, S. Jaimungal, *A mean-field game approach to equilibrium pricing in solar renewable energy certificate markets*, Mathematical Finance, **32**(2022), pp. 779-824. doi: 10.1111/mafi.12345
[30] F. Liu, M. Yu, B. Wei, F. Xiao, *Mean field games of energy storage devices: A finite difference analysis*, Asian Journal of Control, 2023. Published online. doi: 10.1002/asjc.3161
[31] I.T. Michailidis, S. Baldi, M.F. Pichler, E.B., Kosmatopoulos, J.R. Santiago, Proactive control for solar energy exploitation: A german high-inertia building case study, Applied Energy, 155(2015), 409-420. doi: 10.1016/j.apenergy.2015.06.033
[32] H. Yoshioka, Y. Yoshioka, Assessing fluctuations of long-memory water environmental indicators based on the robustified dynamic Orlicz risk, Chaos, Solitons & Fractals, 180(2024), 114336. doi: 10.1016/j.chaos.2023.114336

# APPENDICES

## A. Proof of Proposition 1

We prove **Proposition 1** for generic $d$-dimensional controlled diffusion process ($d \in \mathbb{N}$). The model in the main text is a special case of that in this appendix.

A controlled diffusion process at time $t$ is represented as $\mathbf{X}_t = (X_{1,t}, X_{2,t}, ..., X_{d,t})$, which is governed by the Itô's SDE

$$d\mathbf{X}_t = a(t, \mathbf{X}_t, u_t)dt + b(t, \mathbf{X}_t, u_t)d\mathbf{B}_t, \ t > 0 \tag{25}$$

with an initial condition $\mathbf{X}_0$, where $\mathbf{B}_t = (B_{1,t}, B_{2,t}, ..., B_{d,t})$ represents a $d$-dimensional standard Brownian motion, $a, b : \mathbb{R}^{d+1} \times A \to \mathbb{R}$ with a bounded (and possibly state-dependent) set $A$ are sufficiently smooth drift and diffusion coefficients. The $i$ th element of each vector is denoted with the subscript $i$, e.g., $a_i$. We assume that correlations among each element of $\mathbf{B}_t$ are 0 for simplicity, which suffices to our purpose. The process $u = (u_t)_{t \geq 0}$ is the control variable that is assumed to be measurable with respect to $\mathcal{F}$ as common in classical control problems. The admissible set of $u$ such that it is uniformly bounded, measurable with respect to $\mathcal{F}$, and the SDE (25) admits a unique path-wise continuous solution, is expressed as $\mathcal{A}$.

As in the main text, model uncertainty is represented by a distortion of the drift based on the robust control approach [17]. We set the drifted Brownian motion $\mathbf{W} = (\mathbf{W}_t)_{t \geq 0}$ by

$$dW_{i,t} = dB_{i,t} - \phi_{i,t}dt \ (i = 1, 2, 3, ..., d), \tag{26}$$

where $\phi = (\phi_{1,t}, \phi_{2,t}, ..., \phi_{d,t})_{t \geq 0}$ is some measurable and adapted process. The probability measure under which $W$ is a $d$-dimensional standard Brownian motion without any correlations among its different elements is denoted as $\mathbb{Q}(\phi)$. The existence of $\mathbb{Q}(\phi)$ is guaranteed if [17]:

$$\mathbb{E}_{\mathbb{P}}\left[\exp\left(\frac{1}{2}\sum_{i=1}^{d}\int_0^t \phi_{i,s}^2 ds\right)\right] < +\infty \text{ for each } t > 0. \tag{27}$$

The collection of real-valued measurable processes $\phi$ satisfying (27) is denoted as $\mathcal{B}$. The unit-time relative entropy $c(\phi)$, the Kullback–Leibler divergence, measuring the difference between $\mathbb{P}$ and $\mathbb{Q}(\phi)$ at time $t$ is calculated as $\frac{1}{2}\sum_{i=1}^{d}\phi_{i,t}^2$.

The Orlicz risk is set for a small time increment $k > 0$ at time $t$ as

$$\inf_{u \in \mathcal{A}} \sup_{\phi \in \mathcal{B}} \mathbb{E}_{\mathbb{Q}(\phi)}\left[\Phi\left(\frac{\Psi_{t+k} + \int_t^{t+k} D(s, u_s, \mathbf{X}_s)ds}{\Psi_t}\right) - \sum_{i=1}^{d}\int_t^{t+k}\frac{\phi_{i,s}^2}{2\eta}ds \bigg| \mathcal{F}_t\right] = 1 \tag{28}$$

with a non-negative measurable coefficient $D(s, u_s, \mathbf{X}_s)$. We seek for a Markovian recursive utility having the form $\Psi_t = \Psi(t, \mathbf{X}_t)$ (with an abuse of notations) that is assumed to be sufficiently smooth.

Given $(u, \phi) \in \mathcal{A} \times \mathcal{B}$, the classical Itô's formula for diffusion processes leads to

$$\begin{aligned}
&\Psi(t+k, \mathbf{X}_{t+k}) \\
&= \Psi(t, \mathbf{X}_t) + \int_t^{t+k}\frac{\partial \Psi}{\partial t}(s, \mathbf{X}_s)ds \\
&\quad + \sum_{i=1}^{d}\int_t^{t+k}\left(a_i(s, \mathbf{X}_s, u_s) + \phi_{i,s}b_i(s, \mathbf{X}_s, u_s)\right)\frac{\partial \Psi}{\partial x_i}(s, \mathbf{X}_s)ds \\
&\quad + \sum_{i=1}^{d}\int_t^{t+k}\frac{1}{2}b_i^2(s, \mathbf{X}_s, u_s)\frac{\partial^2 \Psi}{\partial x_i^2}(s, \mathbf{X}_s)ds \\
&\quad + \sum_{i=1}^{d}\int_t^{t+k}\frac{\partial \Psi}{\partial x_i}(s, \mathbf{X}_s)b_i(s, \mathbf{X}_s, u_s)dW_{i,s}
\end{aligned} \tag{29}$$

A Taylor expansion of $\Phi$ with $\kappa$ having a sufficiently small $|\kappa|$ yields

$$\Phi(1+\kappa) = \Phi(1) + \kappa\Phi'(1) + \frac{1}{2}\kappa^2\Phi''(1) + o(\kappa^2) \tag{30}$$

with the Landau symbol $o(\cdot)$.

From (29), if $k > 0$ is small, we consider (30) with $\kappa$ given by

$$\kappa = \frac{1}{\Psi(t,\mathbf{X}_t)} \begin{pmatrix} \int_t^{t+k} \frac{\partial \Psi}{\partial t}(s,\mathbf{X}_s) \mathrm{d}s \\ + \sum_{i=1}^d \int_t^{t+k} \left( a_i(t,\mathbf{X}_s,u_s) + \phi_{i,s} b_i(s,\mathbf{X}_s,u_s) \right) \frac{\partial \Psi}{\partial x_i}(s,\mathbf{X}_s) \mathrm{d}s \\ + \sum_{i=1}^d \int_t^{t+k} \frac{1}{2} b_i^2(s,\mathbf{X}_s,u_s) \frac{\partial^2 \Psi}{\partial x_i^2}(s,\mathbf{X}_s) \mathrm{d}s \\ + \int_t^{t+k} D(s,u_s,\mathbf{X}_s) \mathrm{d}s \\ + \sum_{i=1}^d \int_t^{t+k} \frac{\partial \Psi}{\partial x_i}(s,\mathbf{X}_s) b_i(s,\mathbf{X}_s,u_s) \mathrm{d}W_{i,s} \end{pmatrix} \tag{31}$$

$$:= \kappa_t + \frac{1}{\Psi(t,\mathbf{X}_t)} \sum_{i=1}^d \int_t^{t+k} \frac{\partial \Psi}{\partial x_i}(s,\mathbf{X}_s) b_i(t,\mathbf{X}_s,u_s) \mathrm{d}W_{i,s}$$

By using $\Phi(1) = 1$ and the isometry of Brownian motions, we obtain

$$\mathbb{E}_{\mathbb{Q}(\phi)}\left[ \left( \frac{1}{\Psi(t,\mathbf{X}_t)} \left( \sum_{i=1}^d \int_t^{t+k} \frac{\partial \Psi}{\partial x_i}(s,\mathbf{X}_s) b_i(s,\mathbf{X}_s,u_s) \mathrm{d}W_{i,s} \right) \right)^2 \right]$$
$$= \mathbb{E}_{\mathbb{Q}(\phi)}\left[ \frac{1}{\Psi^2(t,\mathbf{X}_t)} \sum_{i=1}^d \int_t^{t+k} b_i^2(s,\mathbf{X}_s,u_s) \left( \frac{\partial \Psi}{\partial x_i}(s,\mathbf{X}_s) \right)^2 \mathrm{d}s \right]. \tag{32}$$

Then, we arrive at the equality

$$\inf_{u \in \mathcal{A}} \sup_{\phi \in \mathcal{B}} \mathbb{E}_{\mathbb{Q}(\phi)}\left[ \begin{array}{l} \Phi'(1)\kappa_t + \frac{\Phi''(1)}{2\Psi^2(t,\mathbf{X}_t)} \\ \times \sum_{i=1}^d \int_t^{t+k} b_i^2(s,\mathbf{X}_s,u_s) \left( \frac{\partial \Psi}{\partial x_i}(s,\mathbf{X}_s) \right)^2 \mathrm{d}s \\ - \sum_{i=1}^d \int_t^{t+k} \frac{\phi_{i,s}^2}{2\eta} \mathrm{d}s \end{array} \middle| \mathcal{F}_t \right] = 0. \tag{33}$$

Dividing both sides of (33) by $k > 0$ and letting $k \to +0$ yields the HJB equation

$$\inf_{u \in A} \sup_{\phi \in \mathbb{R}^d} \left\{ \Phi'(1)\tilde{\kappa} + \frac{\Phi''(1)}{2\Psi^2} \sum_{i=1}^d b_i^2 \left( \frac{\partial \Psi}{\partial x_i} \right)^2 - \sum_{i=1}^d \frac{\phi_i^2}{2\eta} \right\} = 0 \tag{34}$$

with

$$\tilde{\kappa} = \frac{1}{\Psi} \left( \frac{\partial \Psi}{\partial t} + \sum_{i=1}^d \left( (a_i + \phi_i b_i) \frac{\partial \Psi}{\partial x_i} + \frac{1}{2} b_i^2 \frac{\partial^2 \Psi}{\partial x_i^2} \right) + D(t,u,\mathbf{x}) \right). \tag{35}$$

The quantities inside the expectation right-hand side of the parentheses "{}" of (34) are evaluated at $(t,\mathbf{x})$. Note that we have assumed $\Psi > 0$, with which each term of the HJB equation is well-defined. This HJB equation is subject to some terminal condition at a terminal time $T > 0$. We set the homogenous terminal condition $\Psi(T,\mathbf{x}) = 0$.

As long as $\Psi > 0$, the HJB equation (34) can be written as

$$\inf_{u \in A} \sup_{\phi \in \mathbb{R}^d} \left\{ \Psi \tilde{\kappa} + \frac{\Phi''(1)}{2\Phi'(1)\Psi} \sum_{i=1}^d b_i^2 \left(\frac{\partial \Psi}{\partial x_i}\right)^2 - \Psi \sum_{i=1}^d \frac{\phi_i^2}{2\Phi'(1)\eta} \right\} = 0, \tag{36}$$

which is further rewritten as

$$\inf_{u \in A} \left\{ \begin{array}{l} \dfrac{\partial \Psi}{\partial t} + \sum_{i=1}^d \left( a_i \dfrac{\partial \Psi}{\partial x_i} + \dfrac{1}{2} b_i^2 \dfrac{\partial^2 \Psi}{\partial x_i^2} \right) + D(t, u, \mathbf{x}) \\ + \dfrac{1}{2\Psi} \dfrac{\Phi'(1)^2 \eta + \Phi''(1)}{\Phi'(1)} \sum_{i=1}^d b_i^2 \left(\dfrac{\partial \Psi}{\partial x_i}\right)^2 \end{array} \right\} = 0, \tag{37}$$

or equivalently

$$\inf_{u \in A} \sup_{\phi \in \mathbb{R}^d} \left\{ \begin{array}{l} \dfrac{\partial \Psi}{\partial t} + \sum_{i=1}^d \left( (a_i + \phi_i b_i) \dfrac{\partial \Psi}{\partial x_i} + \dfrac{1}{2} b_i^2 \dfrac{\partial^2 \Psi}{\partial x_i^2} \right) \\ + D(t, u, \mathbf{x}) - \Psi \sum_{i=1}^d \dfrac{\phi_i^2}{2\eta'} \end{array} \right\} = 0 \tag{38}$$

with the net uncertainty-aversion parameter given by $\eta' = \dfrac{\Phi'(1)^2 \eta + \Phi''(1)}{\Phi'(1)} > 0$.

***Remark A1.*** The form (38) of the HJB equation suggests that it is an optimality equation of the following differential game:

$$\Psi(t, \mathbf{X}_t) = \inf_{u \in \mathcal{A}} \sup_{\phi \in \mathcal{B}} \mathbb{E}_{\mathbb{Q}(\phi)} \left[ \int_t^T e^{-\sum_{i=1}^d \int_t^s \frac{\phi_{i,\tau}^2}{2\eta'} d\tau} D(s, u_s, \mathbf{X}_s) ds \,\middle|\, \mathcal{F}_t \right] \tag{39}$$

for $0 \leq t \leq T$. The problem (39) is a stochastic differential game with the control-dependent discount rate $\sum_{i=1}^d \dfrac{\phi_{i,s}^2}{2\eta'}$. Note that the SDE (25) under $\mathbb{Q}(\phi)$ also depends on the control $\phi$ through the transformation (26). See **Appendix C**.

## B. Constrained Viscosity Solutions

We give few remarks concerning weak solutions to the HJB equation (13). This equation should be understood in a viscosity sense as it has a degenerate diffusion coefficient, due to that the solutions possibly become non-smooth such that the partial derivatives are defined only in a weak sense (Crandall et al., 1992). Our HJB equation should be satisfied both inside the domain and along its boundary because there is no information coming from outside the domain and the state variables are a.s. confined in the domain. The notion of constrained viscosity solutions should be applied in this case. Loosely speaking, viscosity subsolution properties are satisfied except along the boundary of $[0,1] \times [0, \bar{Y}]$ while supersolution properties are satisfied both inside the domain and along the boundary of $[0,1] \times [0, \bar{Y}]$.

We do not go deep into constrained viscosity solutions to our HJB equation because of the unique nonlinearity emerging as the coefficient $\dfrac{1}{\Psi}\left(\dfrac{\partial \Psi}{\partial x}\right)^2$. Its quadratic dependence on $\dfrac{\partial \Psi}{\partial x}$ is outside the scope of the classical comparison principle (Tran, 2021). Further, this term may become unbounded as $\Psi$ gets closer to 0, while it is not always so if the coefficient $\left(\dfrac{\partial \Psi}{\partial x}\right)^2$ decays not slower than the speed of the convergence of $\Psi$ toward 0. Our computational results suggested that the latter case happens near the terminal time $t = T$ in the demonstrative computational example of **Section 4**. Indeed, due to enforcing the terminal condition $\Psi = 0$ at the terminal time $t = T$, we can infer $\Psi = \dfrac{\partial \Psi}{\partial x} = 0$ with which the term $\dfrac{1}{\Psi}\left(\dfrac{\partial \Psi}{\partial x}\right)^2$ is undefined at $t = T$. The computational results suggest that the term $\dfrac{1}{\Psi}\left(\dfrac{\partial \Psi}{\partial x}\right)^2$ is small at $t < T$ near $T$. This point needs to be elaborated in the future to better understand the Orlicz risk and associated HJB equation. The boundary treatment in the numerical computation needs to be sophisticated if one considers rigorous convergence in the viscosity sense, which was not addressed in this study.

## C. Proof of Proposition 2

The proof directly tracks Theorem 6.1 of Øksendal and Sulem [6]. Indeed, the problem (17) is a zero-sum stochastic differential game with a control-dependent discount factor. Such a discount factor has not been studied in [6], while no technical difficulties arise in our case due to the regularity assumption of $\Psi$ in $\Omega$ and the homogenous terminal condition $\Psi|_{t=T} = 0$. The same reasoning applies to the more general controlled processes discussed in **Appendix A**.

## D. An Exactly-solvable Case

We present an elementary example where the new HJB equation (34) admits a closed-form solution. We assume the Cox-Ingersoll-Ross model as a famous SDE (Cox et al., 1985):

$$dX_t = (a - rX_t)dt + \sigma\sqrt{rX_t}\,dB_t,\ t > 0,\ X_0 > 0 \tag{40}$$

with $a, r, \sigma > 0$. We assume that there is no control, i.e., only the null control is admissible $\mathcal{A} = \{0\}$, and $D = 0$. The corresponding HJB equation based on the Orlicz risk reads

$$\frac{\partial \Psi}{\partial t} + (a - rx)\frac{\partial \Psi}{\partial x_i} + \frac{1}{2}\sigma^2 rx \frac{\partial^2 \Psi}{\partial x^2} + \frac{\eta'}{2\Psi}\sigma^2 rx \left(\frac{\partial \Psi}{\partial x}\right)^2 = 0,\ t < T,\ x \geq 0, \tag{41}$$

where we assume the terminal condition

$$\Psi(T, x) = e^{px},\ x > 0 \tag{42}$$

with some $p \in \mathbb{R}$ at a terminal time $T > 0$. This problem corresponds to the evaluation of the exponential utility $\mathbb{E}\left[e^{pX_T}\right]$ under uncertainty.

We exactly solve the problem (41)-(42). Guessing the solution of the exponential form

$$\Psi(t, x) = e^{\alpha_t x + \beta_t},\ t \leq T,\ x > 0 \tag{43}$$

with time-dependent parameters $\alpha_t$, $\beta_t$ and substituting (43) into (41) yield the ordinary differential equations for $t < T$:

$$\frac{d\alpha_t}{dt} - r\alpha_t + r\frac{\sigma^2(1+\eta')}{2}(\alpha_t)^2 = 0 \tag{44}$$

and

$$\frac{d\beta_t}{dt} + a\alpha_t = 0 \tag{45}$$

subject to the terminal conditions

$$\alpha_T = p,\ \beta_T = 0. \tag{46}$$

The equation (44) is a logistic-type equation which is exactly solved for $t \leq T$ as

$$\alpha_t = \frac{1}{\frac{\sigma^2(1+\eta')}{2} + \left(\frac{1}{p} - \frac{\sigma^2(1+\eta')}{2}\right)e^{r(T-t)}}. \tag{47}$$

The other equation (45) can be solved by a direct integration as long as the denominator of (47) remains positive.

*E. On Numerical Solutions*

We show that the numerical solution generated by the proposed finite difference method is positive as long as $\Delta t > 0$ is small for each given $\Delta x, \Delta y > 0$. The proof follows based on an induction argument for the common time-explicit monotone discretization for the HJB and related equations (i.e., discretization without considering $I_3$) (Oberman, 2006) combined with the non-negativity of $I_3$ clearly demonstrated in (24) as long as $\Psi_{i+1,j,k} \geq 0$. This non-negativity is thanks to be employing the Godunov-like discretization that exploits the non-negativity of $\bar{p}^2$ under the discretization [22].

We finally report here that the discretization method of $I_3$ does not significantly affect the numerical solutions. **Fig. E1** compares the computed $u^*$ with $\eta = 1$ at 171.5 (day) for Kanazawa for the two different discretization methods for $\bar{p}^2$: the one used in the main text (24) and the other is the following central one

$$\bar{p}^2 = \begin{cases} p_R^2 & (j = 0) \\ \left(\dfrac{p_R + p_L}{2}\right)^2 & (0 < j < N_x) \\ p_L^2 & (j = N_x) \end{cases}. \tag{48}$$

**Fig. E1** demonstrates that there is no significant difference in practice among the numerical solutions generated by the different schemes even under the relatively large model uncertainty. The comparison here is heuristic but supports the use of the discretization (24) because it is comparable to the standard central one (48).

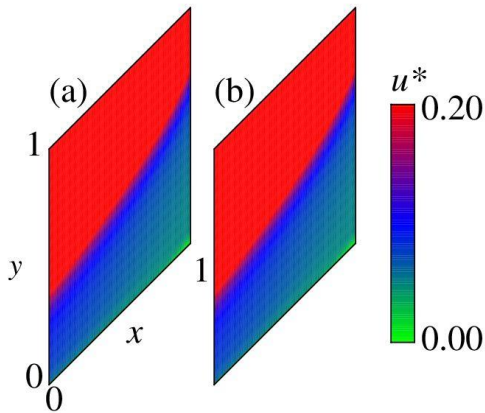

**Fig. E1.** $u^*$ at 171.5 (day): (a) Central discretization and (b) the discretization used in the main text.

We finally demonstrate a computational result on the worst-case uncertainty $\phi^*$ of (14). **Fig. E2** shows $\phi^*$ at for Kanazawa for different values of $\eta$ (hence the net uncertainty aversion $\eta'$) at 171.5 (day). The magnitude of the worst-case uncertainty $\phi^*$ becomes visibly larger as the uncertainty aversion increases as expected. In particular, up to the moderate values of $\eta(=1)$, the computational results suggest that $\phi^*$ is negative (resp., positive) for relatively high (resp., low) energy storage. Considering **Fig. 6**, this is understood as follows; for small energy storage, it is optimal to expect that the cloudy sky will more persist. By contrast, the opposite attitude becomes optimal for large energy storage as the clearer sky more likely to increase the storage but with a larger penalization of not fully discharging energy (i.e., $U - u > 0$) to maintain the storage level. For the larger value of $\eta(=3)$, $\phi^*$ is almost negative due to the excessively high uncertainty that enforces the decision-maker to be more pessimistic about the cloudiness.

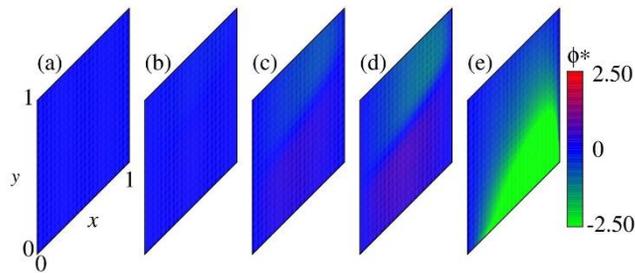

**Fig. E2.** $\phi^*$ at 171.5 (day). $\eta$ is: (a) 0.01, (b) 0.1, (c) 0.5, (d) 1, (e) 3.

**REFERENCES FOR APPENDICES**

J.C. Cox, J.E. Ingersoll, S.A. Ross, *A theory of the term structure of interest rates*, Econometrica, **53**(1985), pp. 385-407. doi: 10.2307/1911242

M.G. Crandall, H. Ishii, P.L. Lions, *User's guide to viscosity solutions of second order partial differential equations*, Bulletin of the American mathematical society, **27**(1992), pp. 1-67. doi: 10.1090/S0273-0979-1992-00266-5

M.A. Katsoulakis, *Viscosity solutions of second order fully nonlinear elliptic equations with state constraints*. Indiana University Mathematics Journal, 1994, pp. 493-519. https://www.jstor.org/stable/24898087

Tran, H. V. (2021). Hamilton–Jacobi equations: theory and applications (Vol. 213). American Mathematical Soc. USA.

A. M. Oberman, *Convergent difference schemes for degenerate elliptic and parabolic equations: Hamilton-Jacobi equations and free boundary problems*, SIAM Journal on Numerical Analysis, **44**(2006), pp. 879-895. doi: 10.1137/S0036142903435235